\documentstyle[aps,multicol,prl]{revtex}
\begin{document}
\input{epsf}
\draft
\title{Retroactive quantum jumps in a strongly-coupled atom--field system}
\author{H. Mabuchi$^{1,2}$ and H.M. Wiseman$^{1}$}
\address{$^1$Department of Physics, University of Queensland, Queensland 4072 Australia\\
$^2$Norman Bridge Laboratory of Physics 12-33, California Institute of Technology,
Pasadena CA 91125 USA}
\date{July 3, 1998}
\maketitle

\begin{abstract}
We investigate a novel type of conditional dynamic that occurs in
the strongly-driven Jaynes-Cummings system with dissipation.  Extending the work
of Alsing and Carmichael [Quantum Opt. {\bf 3}, 13 (1991)], we present a combined
numerical and analytic study of the Stochastic Master Equation that describes the
system's conditional evolution when the cavity output is continuously observed via
homodyne detection, but atomic spontaneous emission is not monitored at all.  We
find that quantum jumps of the atomic state are induced by its dynamical coupling
to the optical field, in order {\em retroactively} to justify atypical fluctuations
occurring in the homodyne photocurrent.
\end{abstract}

\begin{multicols}{2}

\newcommand{\beq}{\begin{equation}} 
\newcommand{\eeq}{\end{equation}}
\newcommand{\bqa}{\begin{eqnarray}} 
\newcommand{\eqa}{\end{eqnarray}}
\newcommand{\nn}{\nonumber} 
\newcommand{\erf}[1]{Eq.~(\ref{#1})}
\newcommand{\dg}{^\dagger}
\newcommand{\rt}[1]{\sqrt{#1}\,}
\newcommand{\smallfrac}[2]{\mbox{$\frac{#1}{#2}$}}
\newcommand{\half}{\smallfrac{1}{2}} 
\newcommand{\bra}[1]{\langle{#1}|} 
\newcommand{\ket}[1]{|{#1}\rangle}
\newcommand{\ip}[1]{\langle{#1}\rangle}
\newcommand{\sch}{Schr\"odinger } 
\newcommand{\schs}{Schr\"odinger's }
\newcommand{\hei}{Heisenberg } 
\newcommand{\heis}{Heisenberg's }
\newcommand{\bl}{{\bigl(}}
\newcommand{\br}{{\bigr)}} 
\newcommand{\ito}{It\^o }
\newcommand{\str}{Stratonovich } 
\newcommand{\dbd}[1]{\frac{\partial}{\partial {#1}}}

Quantum trajectory theories \cite{Carm93,WisMil93a,Dum92,Dali92} have
proven to be of paramount importance in contemporary quantum optics.  This is
largely because they provide powerful computational tools for predicting the
correlation functions and optical spectra of systems with many active degrees
of freedom.  However, quantum trajectories have recently begun to play an
equally important role as the essential theoretical basis for describing
{\em conditional evolution} of continuously-observed open quantum systems.

In this Letter, we use the Stochastic Master Equation (SME) formalism developed
in Ref.~\cite{WisMil93a} to reveal a new type of conditional-dynamical
phenomenon that
occurs in a strongly-coupled open quantum system under {\em partial} observation.
We call this phenomenon {\em retroactive quantum jumps}.  We believe that this
work represents the first use of a measurement-based SME in analyzing a dynamical
behavior specific to partially-observed systems.  Our analysis also illustrates
the utility of more traditional approaches, in particular the use of the
Glauber-Sudarshan $P\left(\alpha\right)$-function \cite{Gar85}, in deriving
simplified conditional evolution equations that retain all the essential features
of a system's quantum dissipative dynamics.

The particular physical system we have studied is the driven Jaynes-Cummings model
\cite{JCM} with dissipation \cite{Carm93,Berm94}. This consists of a two-level
atom resonantly coupled to a resonantly driven cavity mode.  The two
output channels for this system are atomic spontaneous emission into non-cavity
optical modes (at an overall rate of $2\gamma_\perp$), 
and leakage of photons
from the cavity mode through an output-coupling mirror (at rate $2\kappa$).
We focus on the strong atom-cavity coupling limit 
$g \agt \kappa,\gamma_\perp $, and also assume a strong driving field
$E$.  The optical input-output relations for an atom-cavity
system of this type have been experimentally investigated in
Refs.~\cite{Mabu96b,Hood98}.

In a frame rotating at the driving laser frequency, 
the unconditional master equation
is
\beq \label{me1}
\dot\rho = [g(a\dg \sigma - \sigma\dg a)-iEy,\rho ]
+ 2\kappa {\cal D}[a] \rho + 2\gamma_{\perp}{\cal 
D}[\sigma] \rho.
\eeq
Here, for arbitrary operators $A$ and $B$, ${\cal D}[A]B=A\dg B A - 
\half\{A\dg A ,B\}$, and 
$y\equiv-ia+ia\dg$ is the phase quadrature of the field (so that 
$x\equiv a+a\dg$ is the amplitude quadrature), and $\sigma = 
\ket{g}\bra{e}$ is the lowering operator for the atom.

A lot of insight can be gained into this problem by considering the 
corresponding classical equations of motion \cite{AlsCar91}. 
This is done by using the master equation
(\ref{me1}) to calculate the time derivatives  
of the variables 
\beq
\alpha = \ip{a}\;,\;\;s = 
\ip{\sigma}\;,\;\;w=\ip{\sigma_{z}}=\ip{[\sigma\dg,\sigma]},
\eeq
then factorizing all field-atom operator 
products. 
If we ignore spontaneous emission by setting $\gamma_{\perp}=0$, 
we find that the atom will remain in a pure state with 
$w^{2}+4|s|^{2}=1$. Then for $2E>g$ this system  has 
just two fixed points \cite{AlsCar91}
\beq
\alpha_{\pm}^{\rm fix} = \frac{E+gs^{\rm fix}_{\pm}}{\kappa}\;,\;\;
s_{\pm}^{\rm fix} =  - \frac{g}{4E} \mp 
i\left[{\frac{1}{4}-\left(\frac{g}{4E}\right)^{2}}\right]^{1/2}
\eeq
with $w^{\rm fix}=0$. That is, the phase of the field is correlated with the 
state of the atom (which is fully polarized).

\begin{figure}[tb]
\epsfxsize=3.5in
\centerline{\epsfbox{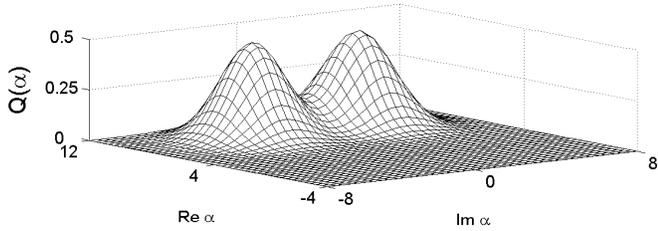}}
\caption{\narrowtext Steady-state $Q(\alpha)$-function for the master
equation (\ref{me1}), with parameters as in Fig.~\ref{fig:SME}.}
\label{fig:Qss}
\end{figure}

In the high driving limit $E\gg g$ (which can be quite realistic), 
these expressions simplify and the two fixed points correspond to
orthogonal quantum states:
\beq
\ket{\psi^{\rm fix}_{\pm}} = \ket{\alpha_{\pm}^{\rm fix}}\ket{\pm}
\equiv \ket{\alpha_{\pm}^{\rm fix}} 2^{-1/2}\left[ \ket{g}\mp 
i\ket{e}\right],
\eeq
where $\ket{\alpha_{\pm}^{\rm fix}}$ is the coherent state with amplitude
\beq \label{newfix}
\alpha_{\pm}^{\rm fix} = E/\kappa \mp i(g/2\kappa)  \equiv \bar\alpha 
\mp i(g/2\kappa).
\eeq
The purity of the atomic state is not preserved if we put back 
spontaneous emission. Nevertheless, if $\gamma_{\perp}$ is small, 
then  the density operator will tend towards an 
equal mixture of the two states $\ket{\psi^{\rm fix}_{\pm}}$ 
\cite{AlsCar91}. We have 
confirmed this by numerically finding the stationary solution of 
\erf{me1}, which has a bimodal $Q\left(\alpha\right)$-function as shown  in
Fig.~\ref{fig:Qss}.  

Our aim is to elucidate the
quantum dissipative dynamics that generate this bimodal distribution, 
in particular the formal mechanisms that enforce correlations
between atomic state and optical phase when the system is subjected to
partial (but continuous) observation.

\begin{figure}[h]
\epsfxsize=3.5in
\centerline{\epsfbox{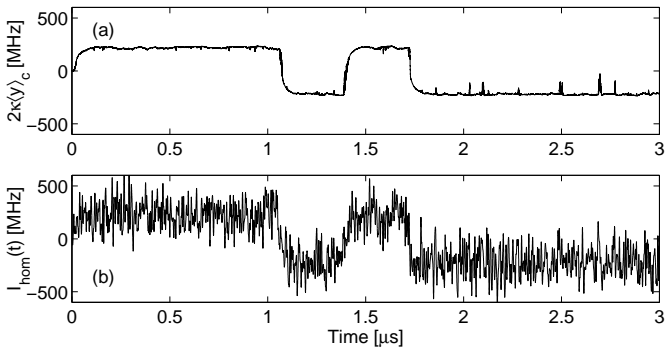}}
\vbox{\vspace{0.1in}}
\caption{\narrowtext Simulation of the master equation 
(\ref{me1}) with the stochastic 
term~\erf{hmt} added (see text for parameters).  (a) conditional
$\ip{y}$; (b) homodyne photocurrent as in \erf{hompc}.}
\label{fig:SME}
\end{figure}

To achieve this aim we first simplify \erf{me1} using a method related 
to that of Ref.~\cite{AlsCar91}. We first transform to an interaction 
picture with respect to the Hamiltonian
\beq
H_{0} = ig\bar{\alpha} (\sigma - \sigma\dg) \equiv 
i(\Omega/2)(\sigma-\sigma\dg).
\eeq
In this picture $\sigma(t) = (-i/2)(\mu e^{-i\Omega t} + \mu_{z} - \mu\dg 
e^{i\Omega t})$, where $\mu=\ket{-}\bra{+}$ and $\mu_{z}=[\mu\dg,\mu]$.
Then, assuming that $\Omega$ is much greater than the characteristic 
rates of atomic evolution $\gamma,g,g^{2}/\kappa$, we can make a 
rotating wave approximation for frequency $\Omega$ to derive
\bqa 
\dot\rho &=& - i [Ey + (g/2)
\mu_{z}x,\rho] + 2\kappa{\cal D}[a]\rho \nn \\ 
&& + \,(\gamma_{\perp}/2)\left\{ {\cal D}[\mu] + {\cal D}[\mu_{z}] +{\cal 
D}[\mu\dg]\right\}\rho. \label{ACme}
\eqa
Here the atomic spontaneous emission has been split into three decay 
channels corresponding respectively to the upper, middle and lower 
peaks of the Mollow triplet.

Alsing and Carmichael, who derived a master equation similar to 
\erf{ACme}, showed that a quantum trajectory unraveling
 based on detecting the
three different photon frequencies would force the coupled atom-field
state into a pure state of the form $\ket{\pm}\ket{\alpha}$ 
\cite{AlsCar91}. Here the coherent amplitude $\alpha$ of the field 
evolves smoothly between jumps that change the atomic state.  
Between jumps the
field state is attracted to the fixed point $\alpha_{\pm}^{\rm
fix}$ corresponding to the current atomic state $\vert\pm\rangle$.
If $\gamma_{\perp} \ll \kappa$, then on a long time scale 
these are occupied with equal probability.

While the unraveling based on observation of atomic decays provides 
an intuitive picture of the dynamics, 
high-efficiency frequency-resolved monitoring of atomic fluorescence
is not yet experimentally feasible.
Given the strong correlation between atomic state and optical phase, however,
it should be possible to observe state-changing atomic decays `indirectly'
via  homodyne monitoring of the phase-quadrature of the cavity output.
This would be much easier to implement in the
laboratory.  One would expect to see bistability of the field with 
values $\alpha_{\pm}^{\rm fix}$, with stochastic switching induced 
(according to the intuitive picture outlined
above) by atomic spontaneous emission.  
But from a theoretical perspective, we must now ask how jump-like
behavior could emerge from the evolution equations for a situation
in which no counting or projective measurements are assumed to be made.  In
what sense should we be able to associate observed phase-switching events with
`actual' atomic decays?

 From the theory of Ref.~\cite{WisMil93a}, homodyne monitoring
of the cavity output can be modeled by adding to the master equation the
following nonlinear, stochastic term:
\beq \label{hmt}
\dot\rho_{\rm meas} = -i\rt{2\kappa\eta}\xi(t) \left\{ a\rho - \rho 
a\dg - {\rm Tr}[\rho(a-a\dg)] \right\} \rho.
\eeq
Here the efficiency of the measurement is $\eta$,   
and $\xi(t)$ represents Gaussian white 
noise, to be interpreted in the \ito sense \cite{Gar85}. 
The measured homodyne photocurrent is 
\beq \label{hompc}
I_{\rm hom}(t) = 2\kappa\eta{\rm Tr}[\rho(t) y] + 
\rt{2\kappa\eta}\xi(t).
\eeq

Simulations of the phase-quadrature homodyne photocurrent, using the full
master equation (\ref{me1}) with \erf{hmt} added, were done using 
$\left(g,\kappa,\gamma_\perp\right)$ = $\left(120,40,2.6\right)$ MHz
(where MHz $\equiv 10^6$ $s^{-1}$).  These parameters correspond to the
recent experiment by Hood {\it et al} \cite{Hood98}.  We assume perfect
detection ($\eta=1$) and set $\left(E/\kappa\right)^2=20$.  This is an
intensive numerical problem \cite{note1}, so the simulations were performed
using a parallel C++/MPI code running on (typically) 64 nodes of an SGI/Cray
Origin-2000 supercomputer.  As is clear from Fig.~\ref{fig:SME}, the simulated
homodyne photocurrent is attracted to the values $\pm 2g$, as expected from 
Eqs.~(\ref{newfix}),(\ref{hompc}).  There is some diffusive noise and stochastic
switches occur at random intervals.  

\begin{figure}[tbh]
\epsfxsize=3.5in
\centerline{\epsfbox{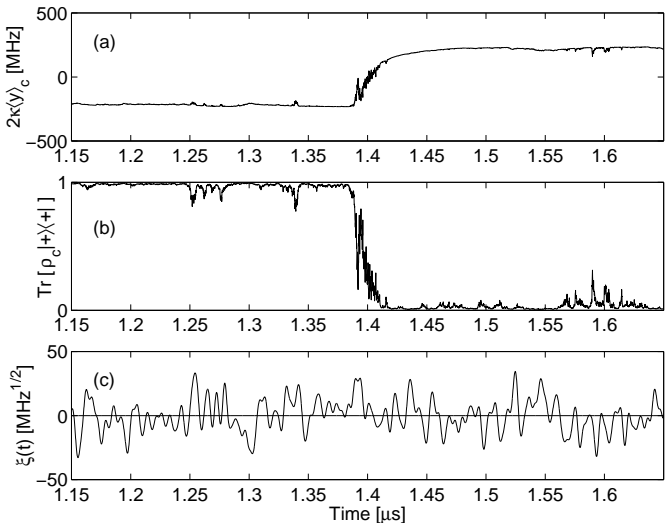}}
\caption{\narrowtext A close-up of  
the simulation in Fig.~2 
showing (a) a switching event in the conditional mean of $y$, (b) the
corresponding jump in atomic state, (c) 
the low-pass filtered noise sequence $\xi(t)$. Note the  
anomalous positively-biased part of the sequence (with a `spike' at 
$t\approx 1.39$)
that caused the jump.
}
\label{fig:rqj}
\end{figure}

 From the simulations, the average rate of switching is $\gamma_\perp/2$, in
agreement with the picture of atomic jumps in Ref.~\cite{AlsCar91}.  Moreover,
the atomic state closely follows the homodyne photocurrent, jumping almost
simultaneously with each phase-switching event (see Fig.~\ref{fig:rqj}).
It must be remembered that {\em there are no explicit jump terms} in the
SME that we have integrated, as we assume no monitoring of the atomic
fluorescence.  Instead, the diffusive noise term $\xi\left(t\right)$, which
arises from the shot-noise fluctuations of the homodyne local oscillator
amplitude, must somehow conspire with the system's intrinsic dynamics
to produce jump-like behavior at a rate determined by the
spontaneous emission parameter $\gamma_\perp$.

In order to understand this `conspiracy' we attempt to solve the 
simplified master equation (\ref{ACme}) with the homodyne measurement 
term (\ref{hmt}) added. 
We use the following ansatz for the coupled atom--field state
\beq
\rho = \sum_{a=\pm}\ket{a}\bra{a} \otimes \int dy\, 
P_{a}(y)\ket{\bar\alpha +iy/2}\bra{\bar\alpha +iy/2} ,
\eeq
where the field states are coherent states, so that $P(y)$ is really 
the Glauber-Sudarshan $P(\alpha)$ function on a line.
Substituting this into \erf{ACme} yields 
\bqa
\dot{P}_{\pm}(y) &=& \left[\dbd{y} (\pm g+\kappa y) + \rt{2\kappa\eta}\xi(t) 
(y-\ip{y})\right]P_{\pm}(y) \nn \\
&& +\, (\gamma_{\perp}/2) \left[ -P_{\pm}(y) + P_{\mp}(y)\right]. \label{Pfe}
\eqa
There is an implicit coupling between  the two distributions in the measurement
terms because 
\beq
\ip{y} = \sum_{a=\pm} \int dy\, y P_{a}(y) 
= \sum_{a=\pm} p_{a} \ip{y}_{a},
\eeq
where $p_{a} = \int dy\, P_{a}(y) = {\rm Tr}[\rho\ket{a}\bra{a}]$.

The effect of the first term in \erf{Pfe} is to drive the field towards the 
semiclassical fixed point $y^{\rm fix}_{\pm} = \mp g/\kappa$, as 
in \erf{newfix}. The second (measurement) term tries to localize the 
distribution at the current mean $\ip{y}$. The final (spontaneous 
emission) term drives the system towards an equal 
mixture of the two atomic states by locally transferring probability
between $P_+$ and $P_-$ at each point $y$. It is the tension between
these three processes (correlation of atomic state with field phase,
localization of the field phase, and destruction of correlations)
which gives rise to
the discrete switching events between
\begin{figure}[h]
\epsfxsize=3.5in
\centerline{\epsfbox{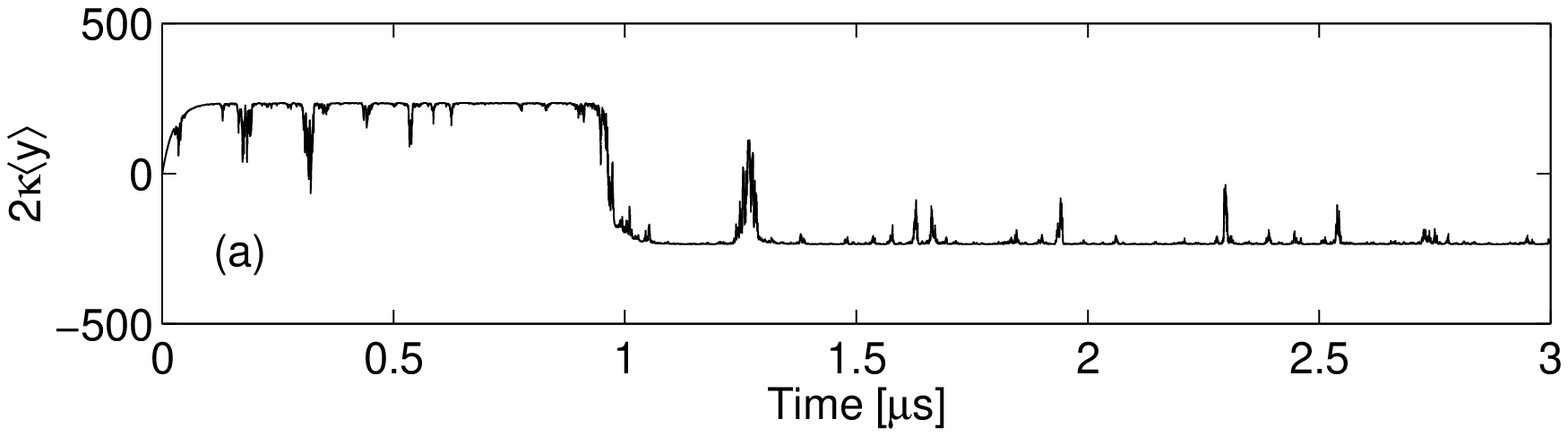}}
\epsfxsize=3.5in
\centerline{\epsfbox{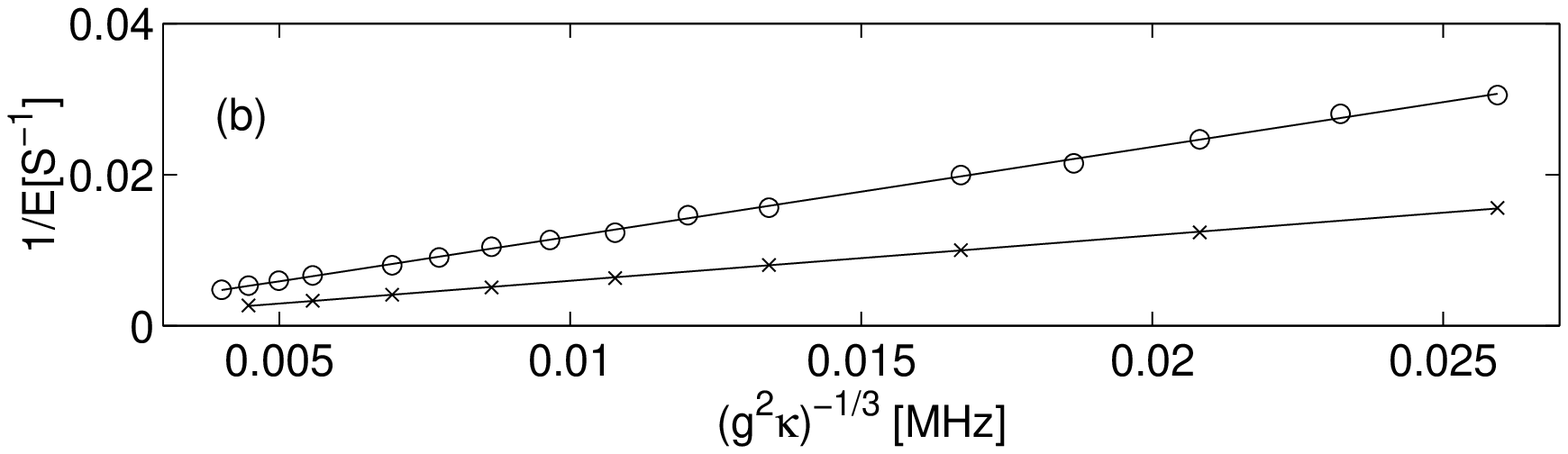}}
\caption{\narrowtext (a) Simulation of the simplified equation~\erf{Pfe}
with parameters as in Fig.~\ref{fig:SME}, (b) scaling of $1/{\rm E}[1/S]$
with $g/\kappa$ fixed ($\eta=1$, $\gamma_\perp=1.3$ MHz for $\circ$'s and
$0.65$ MHz for $\times$'s).}
\label{fig:Pfesim}
\end{figure}
\noindent bistable fixed points. 
The dynamics of \erf{Pfe} is simulated in Fig.~\ref{fig:Pfesim} 
with the same parameter values used with the full SME.  Note that
these simulations were computationally far less demanding than those of
the full SME. The plots in the figure could easily be generated
on a PC \cite{note2}.  We find little difference between the simulated
photocurrents of~\erf{Pfe} and the full SME \cite{note3}.

We can now use the simplified dynamics of \erf{Pfe} to understand how homodyne
detection can cause quantum jumps.  The probability for the atom to be in the
upper state $\ket{+}$ is $p_{+} = \int dy\, P_{+}(y).$ 
 From \erf{Pfe} this obeys
\beq \label{dpu}
\dot{p}_{+} = 
\rt{2\kappa\eta}\xi(t)(p_{+}-p_{+}^{2}) \Delta_{y} - 
\gamma_{\perp}(p_{+} - \half),
\eeq
where $\Delta_{y} \equiv \ip{y}_{+}-\ip{y}_{-} $. Consider the initial 
condition $\ket{\psi}=\ket{\psi_{+}^{\rm fix}}$. 
The damping of the atom (which drives 
$p_{+}$ towards its unconditioned equilibrium value of $\half$) 
immediately produces a 
small component $p_{-}$. 
The field associated with this component will drift towards 
positive values of $y$. Thus $\Delta_{y}$ is negative.  Say $\xi(t)$ then
happens to be generally greater than zero over some short time interval.
Since $0\leq p_{u}\leq 1$, it follows that $p_{+}-p_{+}^{2} \geq 0$.  Thus
the effect of a positive $\xi(t)$ on $p_{+}$ is to decrease it by 
transferring population to state $\ket{-}$.

This implication of Eq.~(\ref{dpu}) may be understood as follows.
A sustained positive trend in $\xi(t)$ indicates a significant 
positive photocurrent fluctuation, 
such as could also take place if the $y$ quadrature
of the field were actually increasing.  Such an
increase in $y$ could be caused by a 
quantum jump of the atom into the $\ket{-}$
state, but would otherwise be unlikely to occur.  The stochastic master equation
agrees with this line of reasoning, but reverses the causality so that occasional
randomly-occurring biases in the photocurrent noise $\xi(t)$ actually {\em cause}
the atom to change its state, as if the jumps are induced retroactively to
justify the atypical photocurrent fluctuations.  Returning to our example, note
that if $\xi(t)$ tends to stay below zero (or fluctuates symmetrically about
zero) then $p_{-}$ will be suppressed, $y$ will stay closed to the fixed point,
and the atom will not have any `reason' to change its state.  This mechanism for
the generation of `retroactive' quantum jumps is confirmed by simulations of
the full SME, as in Fig.~\ref{fig:rqj}.

A quantitative test of our analysis can be made
by considering the non-negative entropy-like variable
\beq
S = p_{+}-p_{+}^{2},
\eeq
which is zero when the atom is in a pure state and $\frac{1}{4}$ when 
it is in a completely mixed state. From \erf{dpu} we can derive the 
following using \ito calculus
\beq
\dbd{t}{\rm E}[\log S] = \kappa\eta {\rm E}\left[( 2S 
-1)\Delta_{y}^{2}\right] 
+\gamma_{\perp}{\rm E}\left[\frac{1-4S}{2S} \right].
\eeq
Here E denotes expectation value with respect to the stochasticity in 
the measurement term (\ref{hmt}), as opposed to the quantum expectation 
values which are denoted as, for example, $\ip{y}_{+}$. Now since we 
expect $S$ to be generally small we can ignore it compared to unity. 
Taking the stationary solution of this equation then gives
\beq \label{oos1}
\gamma_{\perp}{\rm E}[1/(2S)] = \kappa\eta {\rm E}[\Delta_{y}^{2}].
\eeq

To estimate ${\rm E}[\Delta_{y}^{2}]$ we use the fact that the system
stays close to $\ket{\psi_{\pm}^{\rm fix}}$ most of the time. 
Suppose it starts in state $\ket{\psi_{+}^{\rm fix}}$ so that 
$y_{+}=y_{+}^{\rm fix} = -g/\kappa$. 
Then the spontaneous  emission generates
probability at a rate $\gamma_{\perp}/2$ 
for the atom to be in the state
$\ket{-}$.  The associated field $y_{-}$ will drift towards $y_{-}^{\rm fix}$ 
and for short times $t\ll \kappa$ can be approximated by
$y_{-}(t)=-g/\kappa + 2gt$.  This will persist only until the photocurrent
signal it would have generated can be distinguished reliably from the
photocurrent signal generated by the field $y_{+}=y_{+}^{\rm fix}$.  The
integrated difference between the two signals over a time $\tau$ is, from
\erf{hompc}, $\kappa\eta g \tau^{2}$.  The rms-noise in the signal is, again
from \erf{hompc}, $\sqrt{\kappa\eta\tau}$.  According to our explanation for
the retroactive quantum jumps, the atom must decide which state to be in 
at the time $\tau$ such that the
signal and noise are comparable, $\tau \sim (\kappa\eta g^{2})^{-1/3}$. 
It will then (most likely) decide to remain in state $\ket{+}$, and 
the process `repeats' (it is actually  
continuous). The
average of $(y_{+}-y_{-})^{2}$ up to time $\tau$ is easily evaluated to be
$\sim (g/\kappa\eta)^{2/3}$.  Substituting this into \erf{oos1} gives
\beq \label{oos2}
\frac{1}{{\rm E}[S^{-1}]} \sim \frac{\gamma_{\perp}}{ 
2 g^{2/3}(\kappa\eta)^{1/3}}.
\eeq
This formula is valid for $ g \eta^{1/2}\agt \kappa$ and $\gamma_{\perp} \ll
g^{2/3}(\kappa\eta)^{1/3} $. 

The scaling of $1/{\rm E}[1/S]$ with the dynamical parameters of the system was
tested using simulations of \erf{Pfe}.  The results, shown in 
Fig.~\ref{fig:rqj}b,
are in excellent agreement with the prediction within its regime of validity. 
It is interesting that the atomic entropy $\sim 1/{\rm E}[S^{-1}]$ increases 
very slowly ($\sim \eta^{-1/3}$) with decreasing homodyne detection efficiency 
$\eta$. By contrast, detecting the atom's fluorescence as in Ref.~\cite{AlsCar91}
would give an entropy $\sim 1-\eta$.

These results demonstrate that we do understand how the `quantum diffusion'
caused by homodyne monitoring of the cavity can induce `quantum jumps'
of the (unmonitored) atom.  It should be noted that the master equation
for radiative decay does {\em not} imply that the atom  has any intrinsic
preference to undergo jump-like behavior.  As discussed in Ref.~\cite{WisMil93c}
for example, homodyne monitoring of the atom's fluorescence would cause it to
undergo diffusive evolution.  The atomic jumps we have investigated above are
truly a {\em dynamical} consequence of the strong correlation between the atomic
state and the phase of the intracavity field, which itself stems from the
Jaynes-Cummings Hamiltonian.  There is no reason to believe that retroactive
quantum jumps are unique to this particular system.  We expect that using the
stochastic master equation technique to investigate partially-observed
strongly-coupled quantum systems will turn up many other examples of this new
phenomenon.

H.\ M.\ would like to acknowledge the Advanced Computing Laboratory of LANL.
H.M.W.~was supported by the Australian Research Council.

\newpage
\end{multicols}

\begin{references}
\bibitem{note2} We integrated the $P$-function evolution on a grid of 128
points.  Care was required in the placing of grid points and 
the discretization of derivatives
because of the pathological absence of diffusive terms in
equation~\erf{Pfe}.
\bibitem{note3} Differences that do exist may be due to 
the fact that the parameters chosen do not satisfy the condition 
$\Omega \gg g^{2}/\kappa$.
\end{references}
\end{document}